\documentclass[preprint2]{aastex}

\newcommand{\starstotal}{$32,129$\,\,}
\newcommand{\varstotal}{$145$\,\,}
\newcommand{\varsknown}{$20$\,\,}
\newcommand{\varsnew}{$125$\,\,}
\newcommand{\nightstotal}{$52$\,\,}

\begin{document}

\title{Variable Star Census in a Perseus Field}

\author{T. Pasternacki, Sz. Csizmadia, J. Cabrera, P. Eigm\"uller\altaffilmark{1}, A. Erikson, T. Fruth, P. von Paris, H. Rauer\altaffilmark{2} , R. Titz}
\affil{}
\affil{Deutsches Zentrum f\"ur Luft- und Raumfahrt, Institut f\"ur Planetenforschung, Rutherfordstra\ss e 2, 12489 Berlin, Germany}
\email{thomas.pasternacki@dlr.de}

\author{J. Eisl\"offel, A. Hatzes}
\affil{Th\"uringer Landessternwarte Tautenburg, Sternwarte 5, 07778 Tautenburg, Germany}

\author{M. Boer, G. Tournois}
\affil{Observatoire de Haute Provence, St. Michel-l'Observatoire 04870, France}

\author{P. Kabath}
\affil{European Southern Observatory, Alonso de C{\'o}rdova 3107, Vitacura, Casilla 19001, Santiago Chile}

\author{P. Hedelt\altaffilmark{3}}
\affil{CNRS, UMR 5804, Laboratoire d'Astrophysique de Bordeaux, 2 rue de l'Observatoire, BP 89, F-33271 Floirac Cedex, France}

\and

\author{H. Voss}
\affil{Universitat de Barcelona, Departament d'Astronomia i Meteorologia, Marti i Franqu\`es 1, E-08028 Barcelona, Spain}

\altaffiltext{1}{Th\"uringer Landessternwarte Tautenburg, Sternwarte 5, 07778 Tautenburg, Germany}
\altaffiltext{2}{Technische Universit\"at Berlin, Zentrum f\"ur Astronomie und Astrophysik, Hardenbergstra\ss e 36, 10623 Berlin, Germany}
\altaffiltext{3}{Observatoire Aquitain des Sciences de l'Univers,
2 rue de l'Observatoire, BP 89, F-33271 Floirac Cedex, France}

\begin{abstract}
The Berlin Exoplanet Search Telescope (BEST) is a small-aperture, wide-field telescope dedicated to time-series photometric observations. During an initial commissioning phase at the Th\"uringer Landessternwarte Tautenburg (TLS), Germany, and subsequent operations at the Observatoire de Haute-Provence (OHP), France, a 10-square-degree circumpolar field close to the galactic plane centered at $(\alpha, \delta)= (02^h\,39^m\,23^s,\,+52^\circ\,01'\,46'')$ $(J2000.0)$ was observed between August 2001 and December 2006 during 52 nights. From the \starstotal stars observed a subsample of \varstotal stars with clear stellar variability was detected out of which \varsnew are newly identified variable objects. For five bright objects the system parameters were derived by modelling the light curve.
\end{abstract}

\keywords{data analysis --- stars: variable: general --- binaries:eclipsing --- Cepheids --- $\delta$ Scuti --- techniques: photometric}

\section{Introduction}

The Berlin Exoplanet Search Telescope (BEST) is a small-aperture, wide-field telescope dedicated to time-series photometric observations. At present BEST is located at Observatoire de Haute-Provence, France (OHP) and is remotely controlled from Berlin \citep{raue04, raue10}. From 2001 to 2003 BEST was located at the Th\"uringer Landessternwarte (TLS), Germany, for a commission phase. Primarily built-up to provide ground-based support to the CoRoT space mission \citep{bagl06}, BEST offers the great possibility for detection and examination of new variable stars due to its high-precision stellar photometry.

Based on photometric observations of eclipsing binaries, constraints can be set on their orbital inclination and relative radii. Together with radial velocity measurements it is possible to determine absolute system parameters. Only with this knowledge details of the structure and evolution of the observed system becomes available.

The present paper describes the observations of a selected target field located in Perseus. Observations were done between 2001 and 2003 from the TLS site and between 2005 and 2006 from the OHP site. Previously published results based on BEST observations in support of the CoRoT space mission can be found in \citet{karo07,kaba07,kaba08,raue10}.

\section{Telescope and observations}

BEST is a $f/2.7$ Cassegrain-telescope with an aperture of 19.5\,cm and is equipped with an air cooled  $2048\times 2048$ AP-10 CCD-camera. The field-of-view (FOV) has a size of $3.1^\circ\times 3.1^\circ$ and the resulting pixel scale is $5.5$''/pixel. All observations are obtained in white light leading to a bandpass limited by the quantum efficiency of the detector, which peaks in the red.  A more detailed description of the system can be found in \citet{raue04, raue10}.

This paper deals with the variable star census in a target field located in the constellation of Perseus centered on the coordinates:
\begin{eqnarray*}
 \alpha(J2000.0) & = & 02^h\,39^m\,23^s \\
 \delta(J2000.0) & = & +52^\circ\,01'\,46''.
\end{eqnarray*}

The target field was observed with BEST from the TLS site between 2001 August 15 to 2003 March 13 during 25 nights for a total of 70 hours. After the relocation of BEST to OHP in 2004 the same field was observed between 2005 November 7 and 2006 December 20 during 27 nights for a total of 87 hours. 

The observational sequence consisted of an equal number of images with $40$, $120$ and $240$\,s exposure time, followed by a bias and dark image after each second run. Additionally, flat and bias images were acquired at the beginning of the night. Here we only present data with $240$\,s exposure time, corresponding to 1036 frames in total, 566 from TLS and 473 from OHP. From the total data set we were able to detect and obtain light curves for around \starstotal\,stars within a magnitude range between $11$ to $17$\, mag. The data set is available to the scientific community upon request\footnote{contact by email: thomas.pasternacki@dlr.de}.

\section{Data Processing} \label{sec:processing}

\subsection{Reduction pipeline}

The acquired data are processed with an automated pipeline combining experience from earlier reductions within the BEST project \citep{kaba07, karo07, kaba08, kaba09a, kaba09b, raue10}.

The first step comprises bias, dark and flat-field correction for every frame. In order to achieve  high precision photometry we use image subtraction \citep{alar00} to detect brightness variations close to the noise limit. After shifting every frame to a common image coordinate system, a set of ten frames with the best photometric quality is combined to a reference image. Thereafter, the reference image is transformed to every scientific image by convolution with a space-varying kernel. Then, the scientific images are subtracted from the fitted reference image. Following these steps the resulting frames contain information about brightness variations only. We are now able to perform relative photometry on reference and scientific images and have already accounted for the differential extinction.

A simple unit-weighted aperture photometry is applied to the reference frame and to the subtracted frames in order to determine the brightness variation as a function of time. The aperture radius is chosen by checking the Full-Width-Half-Maximum (FWHM) of stars at the bright and faint end of the magnitude range. A value of 5 pixels was chosen. 

In order to account for night to night variations in weather conditions and atmospheric transparency a zero offset correction is done by calculating and comparing the averaged star brightness in every frame for the 500 stars with lowest standard deviation. Then, the SysRem algorithm \citep{tamu05} was used in order to further reduce systematic effects.

For the purpose of astrometric calibration the positions of the detected stars are matched and transformed to the celestial coordinates of the USNO-A2.0 catalog \citep{mone98} with the help of the routines GRMATCH and GRTRANS \citep{pal06}. Finally, the instrumental magnitudes are corrected to the absolute magnitude system of the stars in the USNO-A2.0 catalog. This is done by adding an averaged magnitude offset of bright stars in the stellar catalog and the data set. Since we are primarily interested in differential photometry in order to detect stellar variability, a full calibration to a standard photometric system has not been performed on the data set.

\subsection{Photometric Quality}

Bad weather conditions and systematic offsets between different nights can affect the photometric accuracy of the BEST system. To assess the photometric quality in single nights the standard deviation $(\sigma_i)$ is calculated for the light curves of every star $i$. We found $22$ out of $52$ nights in which more than $2,000$ stellar light curves have a $\sigma_i$ less than $1$\,\% in the magnitude range between $10.5$ and $13$\,mag. These nights are defined as good photometric nights. The number of stars with $\sigma < 1$\,\% of the whole campaign is around $600$. 

\begin{figure}[ht]
%\plottwo{fig1.eps}{fig1_color.eps}
\plotone{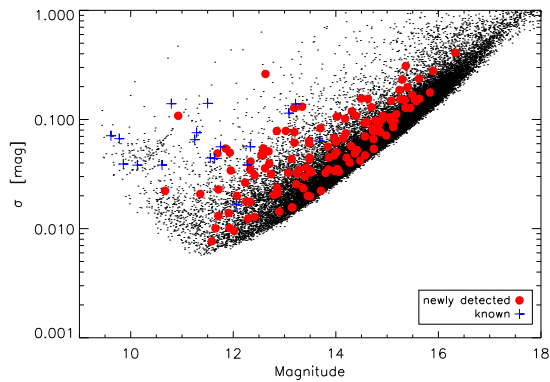}
\caption{Standard deviations $\sigma_i$ vs. magnitude for the whole data set of 52 nights. Known variables are marked with blue crosses and new detections are marked with a red dot. See the electronic edition of the Journal for a color version of this figure.\label{fig:rmsmag}}
\end{figure}

In Fig.~\ref{fig:rmsmag} the standard deviations $\sigma_i$ of the light curves are plotted against the stellar magnitude for the whole data set of 52 nights. Also marked are the known and newly detected variable stars presented in this paper. A comparison between both groups shows that the present survey goes deeper in magnitude then the previous ones. The known variables are more dominant in the bright regime with large $\sigma_i$-values, whereas the newly detected variables are homogeneously distributed in magnitude and $\sigma_i$.

\subsection{Detection}

\begin{figure}[tb]
%\plottwo{fig2.eps}{fig2_color.eps}
\plotone{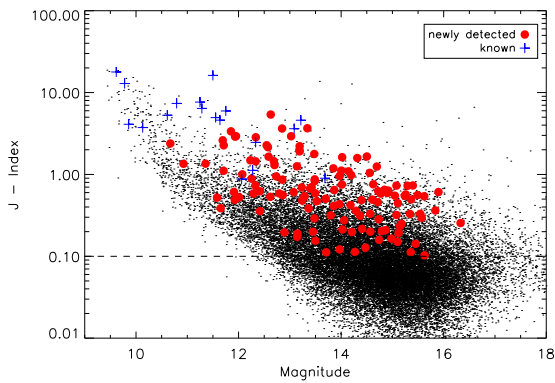}
\caption{Distribution of Stetson's variability index $J$ vs. magnitude shown for all detected stars in the data set. Red dots denote newly detected variable stars with BEST while blue crosses display previously known variable stars. The dashed line marks the lower limit of $J=0.1$  considered for detections. See the electronic edition of the Journal for a color version of this figure.\label{fig:jmag}}
\end{figure}

For identification of potential stellar variability we use variability index J introduced by \citet{stet96} with modified weighting factors according to \citet{zhan03}. The distribution of the calculated $J$-indices versus magnitude is shown in Fig.~\ref{fig:jmag}. Based on the experience acquired in the analysis of previous fields characterized with BEST a limiting value of $J > 0.1$ is chosen to mark potentially variable stars. From the total number of detected stars this limit singles out around  $13,000$ stars which were subsequently examined for periodic variability using the Analysis-of-Variance (AoV) algorithm \citep{sccz96}. Due to the size of the data set at hand, we limited our search to 
periodic signals between 0.1 and 25.0 days. The light curve variations of the objects found in this process are then inspected visually to identify periodic variable stars. The result of this selection will be discussed in the following section.

In particular artificial periods caused by the diurnal cycle lead to many false alarms periods but with no significant presence of stellar variability. Therefore, stars with such periods are excluded in the process if no clear and intrinsic variability can be found by visual inspection.

\section{Results} \label{sec:classification}

\begin{table}[t]
\begin{center}
\caption{Statistical Overview for all detected variable stars. Values in parenthesis state the number of new detections.\label{tab:varstatist}}
\begin{tabular}{ll}
\tableline\tableline
observed stars					&\starstotal\\
stars with $J > 0.1$			&$13,296$\\
\tableline
\textbf{Intrinsic}\\
DCEP								&1 (1)\\
DSCT								&31 (30)\\
RRLYR								&2 (2)\\
\tableline
\textbf{Extrinsic}\\
EA									&24 (21)\\
EB									&6 (6)\\
EW									&45 (42)\\
ACV								&5 (5)\\
ELL								&5 (5)\\
SP									&1 (1)\\
\tableline
other								&25 (12)\\
\tableline
\textbf{Total}\\
variable stars					&\varstotal\,(\varsnew) \\
previously known				&\varsknown\\
\tableline
\end{tabular}
\end{center}
\end{table}

The variable star census in a Perseus target field yielded \varstotal light curves showing clear variable signals, periodic or aperiodic. Thereof \varsnew are newly detected variable objects and \varsknown were previously known. For newly detected variable stars we used a GCVS-based reduced classification mainly separating in intrinsic and extrinsic variable stars \citep{samu09}.

Intrinsic variables are stars whose variability is caused by changing internal stellar properties leading to a change in luminosity, e.g pulsation. We used the subgroups $\delta$~Scuti (DSCT), $\gamma$~Doradus (GDOR), $\delta$~Cephei (DCEP) and RR~Lyrae (RRLYR) stars.

Extrinsic variables show apparent changes in brightness due to external processes like eclipses or rotation. Here we use eclipsing binaries type stars on the one hand, divided into Algol- (EA),  $\beta$- (EB) and W Ursae Majoris-type (EW), and the rotating variables on the other hand, divided into ellipsoidal variables (ELL) and the strong magnetic variables $\alpha$ Canum Venaticorum (ACV). Remaining objects for which no assignment is possible are classified as miscellaneous (MISC).

An overview of the total number for every subtype, split into new and known detections, can be found in Table~\ref{tab:varstatist}. The percentage of variable stars in the whole data set of \starstotal stars amounts to $0.45$\,\% and is of the same order as in other photometric surveys, e.g. ASAS~3 with $0.34$\,\% or OGLE~II with $0.70$\,\% \citep{eyer08}.

The catalog of all detected variable stars in the data set is provided in Table \ref{tab:varcat}. It shows the coordinates, $R$ magnitude, period, epoch, amplitude, and variability type for every detected variable object. Already known objects are marked with flag 'k' and crowded stars are marked with flag 'c'. Crowded in this sense means, that at least two different stars in Table \ref{tab:varcat} show exactly the same period (e.g. F2\_03278 and F2\_03319). These objects are below the angular resolution limit of our CCD camera and have overlapping Point-Spread-Functions (PSFs). The real source of variability for those objects cannot be determined from our data set.

In addition, the reader should be aware of the contamination problem when using the measured amplitudes in Table \ref{tab:varcat}. These are real, if no contaminating objects lie within a radius of $27.5$'' (The pixel scale is $5$''/pixel and the photometric aperture has a radius of $5$\,pixels). Otherwise, if a star is contaminated, the measured amplitude is underestimated. To get rid of this effect the respective system has to be measured with another telescope providing higher angular resolution.

\begin{figure}[tb]
%\plottwo{fig3.eps}{fig3_color.eps}
\plotone{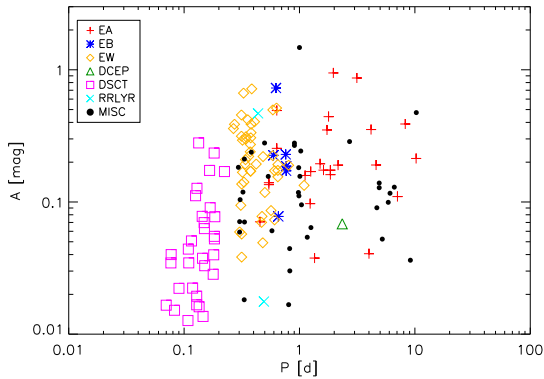}
\caption{Amplitudes of the different kinds of variable stars in the observed field as a function of period. See the electronic edition of the Journal for a color version of this figure.\label{fig:ampper}}
\end{figure}

Fig. \ref{fig:ampper} shows the relation between amplitude, period, and variability type. The period range reaches from $0.07$ to $10$\,days, whereas the amplitude range starts at $0.015$\,mag and ends at $1$\,mag. Note that the shorter the period the smaller the detected amplitudes because of the increase of the signal-to-noise ratio.

In the following paragraphs we discuss the types of the variable stars detected in our data set. Already known variable stars are discussed in Section \ref{sec:knownvars} and the newly detected extrinsic and intrinsic variable stars in Section \ref{sec:extvars} and \ref{sec:intvars}, respectively. 

We further investigated the most interesting and promising objects among the binary detections. Thus, if the whole phase range of an object is covered and if there are enough data points during the eclipses, we are able to derive system parameters for eclipsing binaries. For these cases we modelled the light curves with a binary star light curve modelling code \citep{csiz09}, which assumes Roche-geometry and provides results comparable with the model of \citet{Wils71}. The solutions from this model are optimized via a genetic algorithm as described in \citet{geem01}.

Among the modelled EW-type variables we further distinguish between A-subtype and  W-subtype systems following \citet{binn65} and \citet{csiz04a}. In an A-subtype system the hotter star is the more massive component and it is vice versa for members of a W-subtype system.

\begin{table*}[t]
\tiny
\begin{center}
\caption{Characteristics of the modelled binary systems\label{tab:modelparameters}. The fill-out factors are calculated by $f=\frac{\Omega_{L1}-\Omega}{\Omega_{L2}-\Omega_{L1}}$, where $L_1$ and $L_2$ are the Lagrange-points, and $\Omega$ is the dimensionless surface potential, see e.g. \citet{kopa78}.}
\begin{tabular}{lccccc}
\tableline\tableline
catalog ID								&DV Per				&NSVS 1926064			&NSVS 1913469			&-						&NSVS	4052900		\\
BEST ID									&F2\_02633			&F2\_04262				&F2\_10496				&F2\_10966			&F2\_20731			\\
\tableline
\textbf{Measured parameters}&&&&&\\
Magnitude $R_B$ [mag]				&$13.68$				&$11.50$					&$11.25$					&$12.34$				&$13.21$				\\
Period $P$ [d]							&$1.61427$			&$0.40747$				&$0.42166$				&$0.50111$			&$0.35609$			\\
Epoch $T_0$ [HJD-$2\,452\,195$]	&$0.80714$			&$0.56177$				&$0.61401$				&$0.57373$			&$0.76611$			\\
Amplitude $A$ [mag]					&$0.17$				&$0.40$					&$0.19$					&$0.12$				&$0.40$				\\
color Index $J-K$						&$0.255$				&$0.377$					&$0.345$					&$0.254$				&$0.477$				\\
\tableline
\textbf{System parameters}&&&&&\\
Grav. darkening $g_1$, $g_2$		&$0.32$ (fixed)	&$0.32$ (fixed)		&$0.32$ (fixed)		&$1.0$ (fixed)		&$0.32$ (fixed)	\\
Bol. limb dark. (pri, sec)			&$0.3$ (fixed)		&$0.3$ (fixed)			&$0.3$ (fixed)			&$0.3$ (fixed)		&$0.3$ (fixed)		\\
Albedo $A_1$, $A_2$					&$1.0$ (fixed)		&$0.50$ (fixed)		&$0.50$ (fixed)		&$1.0$ (fixed)		&$0.50$ (fixed)	\\
Mass Ratio $q$							&$1.0$ (fixed)		&$0.244 \pm 0.007$	&$0.25 \pm 0.02$		&$0.27 \pm 0.05$	&$0.23 \pm 0.01$	\\
Temperature $T_1$ [K]				&$6310\pm 120$		&$5130 \pm 180$		&$5630 \pm 180$		&$6660 \pm 380$	&$5764 \pm 481$	\\
Temperature $T_2$ [K]				&$6300$ (fixed)	&$5800$ (fixed)		&$5900$ (fixed)		&$6400$				&$5500$ (fixed)	\\
Inclination $i$ [$^\circ$]			&$85.8\pm 1.2$		&$88.3 \pm 5.1$		&$75.14 \pm 0.09$		&$50.50 \pm 1.54$	&$85.3 \pm 2.2$	\\
Fill-out factor $f_1$  				&$-4.906\pm 0.092$&$f_1=f_2$				&$f_1=f_2$				&$f_1=f_2$			&$f_1=f_2$			\\
Fill-out factor $f_2$  				&$-4.072\pm 0.098$&$0.627 \pm 0.020$	&$0.036 \pm 0.018$	&$0.378 \pm 0.037$&$0.123 \pm 0.037$\\
Limb darkening (pri)					&$0.6$ (fixed)		&$0.01 \pm 0.06$		&$0.25 \pm 0.09$		&$0.42 \pm 0.70$	&$0.80 \pm 0.06$	\\
Limb darkening (sec)					&$0.6$ (fixed)		&$0.53 \pm 0.05$		&$0.49 \pm 0.17$		&$0.49 \pm 0.20$	&$0.50 \pm 0.11$ 	\\
surface potential $\Omega_{pri}$	&$6.4025$			&$7.6460$				&$7.8516$				&$7.2488$			&$f_1=f_2$			\\
surface potential $\Omega_{sec}$	&$5.9515$			&$7.6460$				&$7.8516$				&$7.2488$			&$8.82204$			\\
\tableline
\textbf{Stellar fractional radii}&&&&&\\
\textit{Primary}&&&&&\\
R1 (Pole)								&$0.1845$			&-							&$0.4709$				&$0.4775$			&-						\\
R1 (Side)								&$0.1857$			&$0.5421$				&$0.5089$				&$0.5190$			&$0.5209$			\\
R1 (Back)								&$0.1873$			&$0.5740$				&$0.5334$				&$0.5485$			&$0.5457$			\\
R1 (Point)								&$0.1878$			&-							&-							&-						&-						\\
\\ \textit{Secondary}&&&&&\\
R2 (Pole)								&$0.2012$			&$0.2713$				&$0.2500$				&$0.2697$			&-						\\
R2 (Side)								&$0.2028$			&$0.2863$				&$0.2604$				&$0.2830$			&$0.2581$			\\
R2 (Back)								&$0.2051$			&$0.3497$				&$0.2941$				&$0.3306$			&$0.2945$			\\
R2 (Point)								&$0.2059$			&-							&-							&-						&$0.4799$			\\
\tableline
\end{tabular}
\end{center}
\end{table*}

\subsection{Known Variables} \label{sec:knownvars}

The stars observed with BEST are cross-checked with the Variable Star Index\footnote{http://www.aavso.org/vsx/} (VSX) of the American Association of Variable Star Observers and with the General Catalog of Variable Stars (GCVS) \citep{samu09}. Within the observed target field, in total 26 previously known variable stars could be found according to these catalogs. For \varsknown of these stars we could confirm the previously detected stellar variability. The remaining six cases were too bright to be observed with BEST and consequently were saturated in the acquired data set.

In the variable star catalog (Table~\ref{tab:varcat}) all previously known objects are marked with flag 'k'. For the following 12 long periodic or irregular known variables we are only able to confirm the variability: EF Per, EE Per, EH Per, V0670 Per, V0726 Per, V0727, NSVS J0243375+530503, NSVS J0236346+524133, SAVS 023628+521630, NSVS J0242125+513436, NSV 935, and NSVS J0235501+533958. Confirmation of previously assigned stellar variability classes was only possible for short periodic objects due to the BEST duty cycle. Hereafter we discuss the remaining eight cases of previously known and short periodic variables individually.

\subsubsection{DV Per}

The variability of DV Persei was discovered by \citet{hoff44}. He reported an orbital period of $P=0.807137$\,days, an amplitude of $A_{pri}=0.5$\,mag for the primary and $A_{sec}=0.1$\,mag for the secondary minima, respectively. Minima observations were reported by \citet{kraj06} and \citet{dvor06}, too.

\begin{figure*}[tb]
\centering
\includegraphics[width=0.49\textwidth]{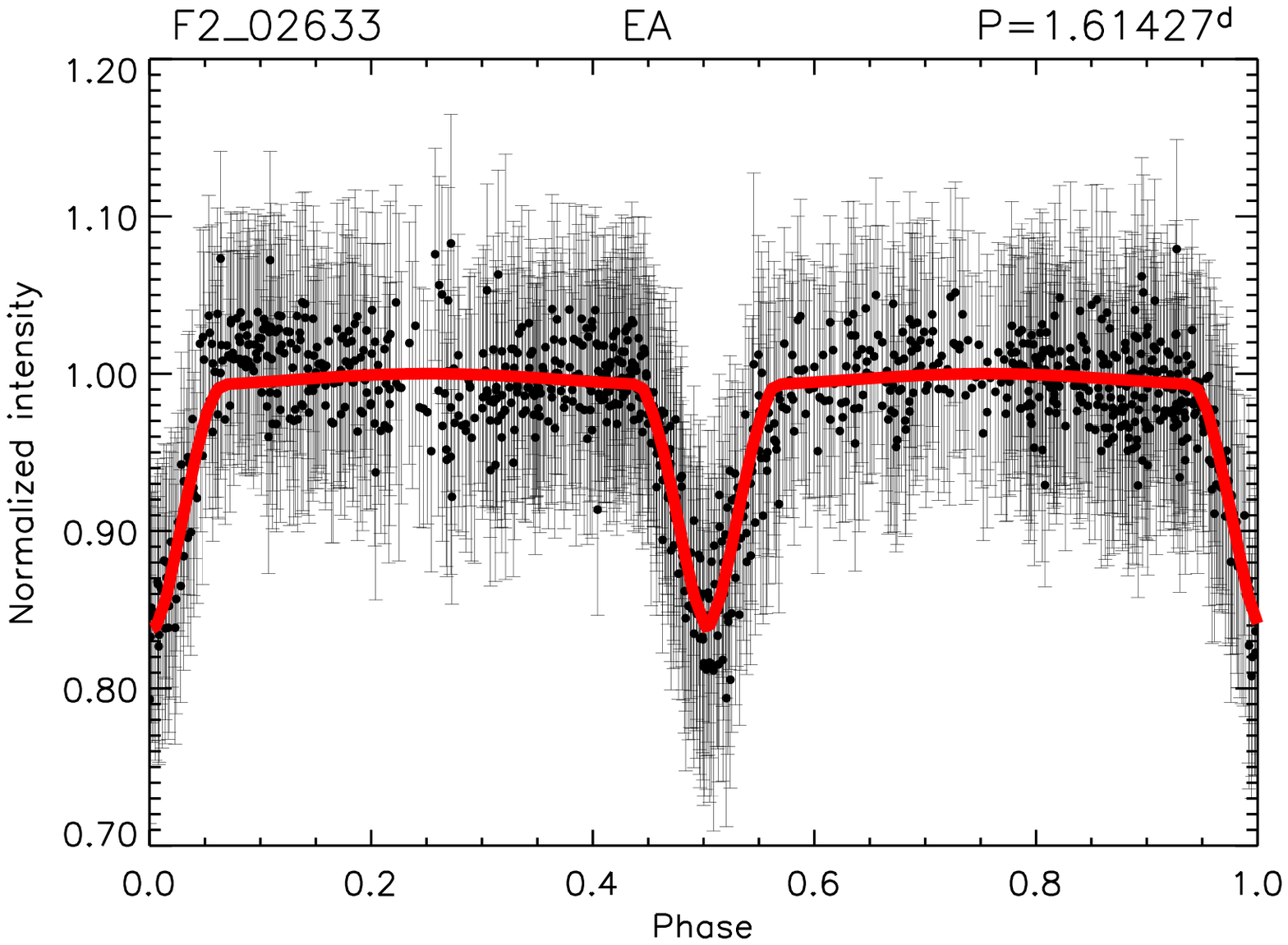}
\includegraphics[width=0.49\textwidth]{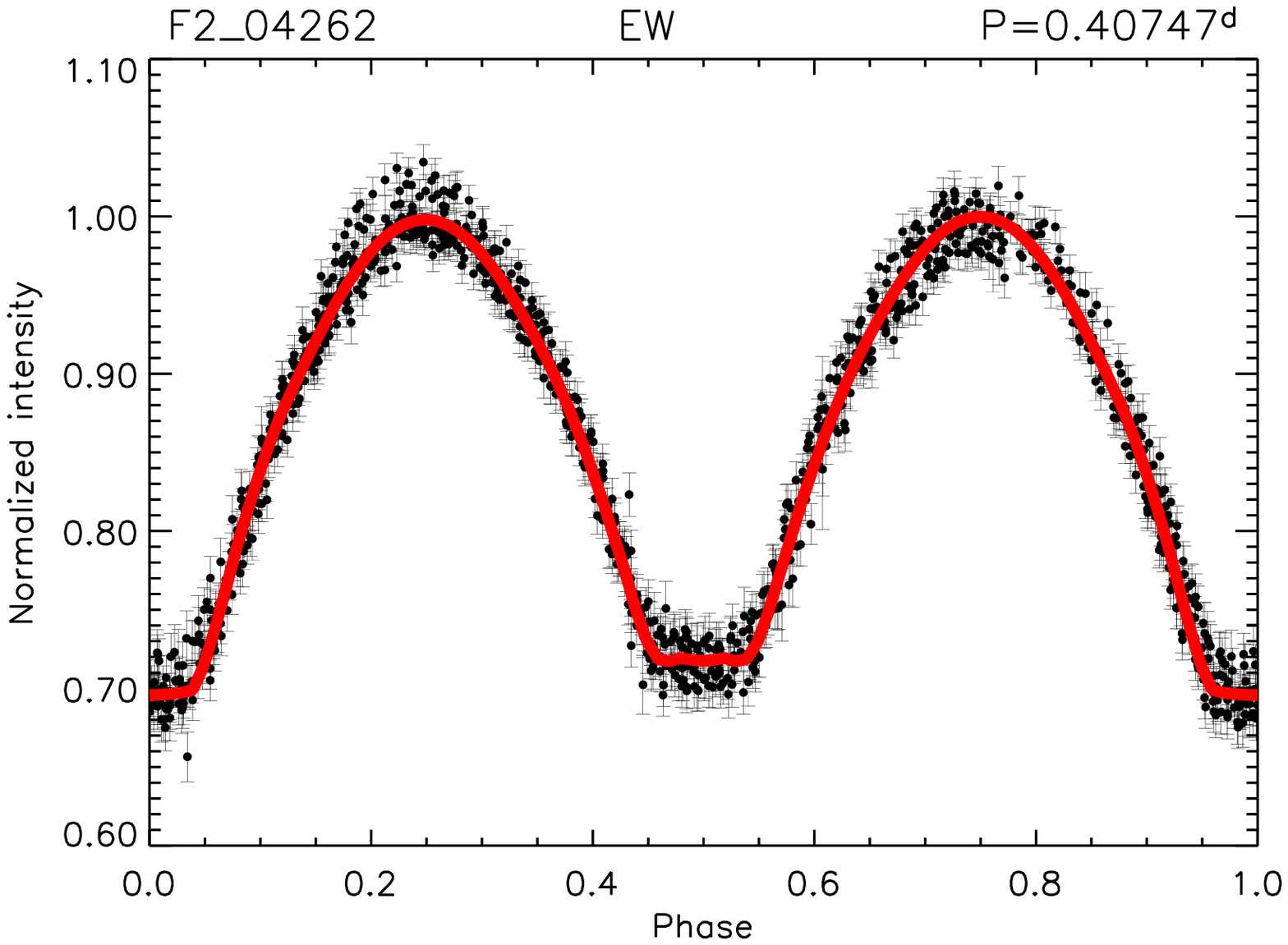}
\includegraphics[width=0.49\textwidth]{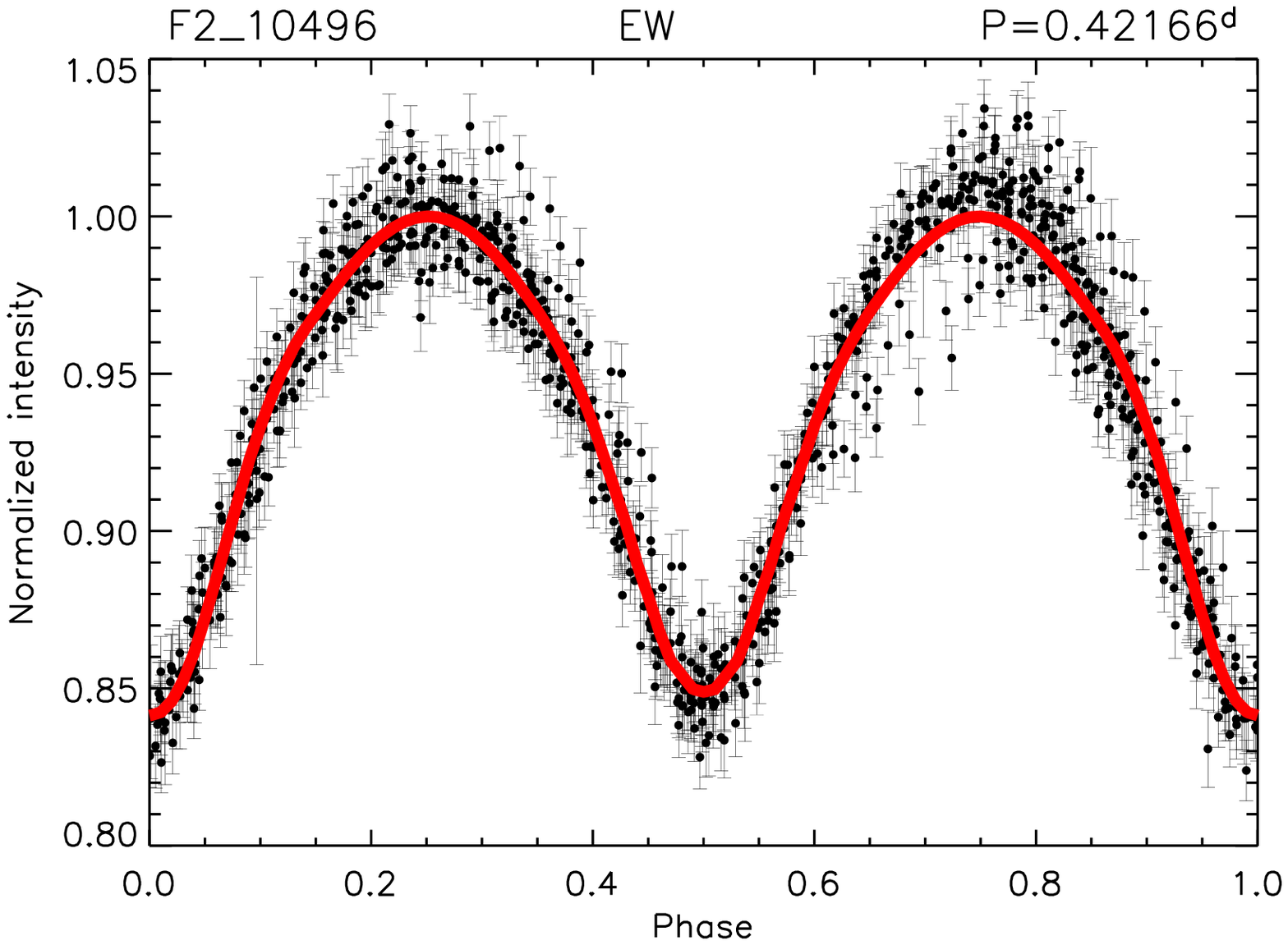}
\includegraphics[width=0.49\textwidth]{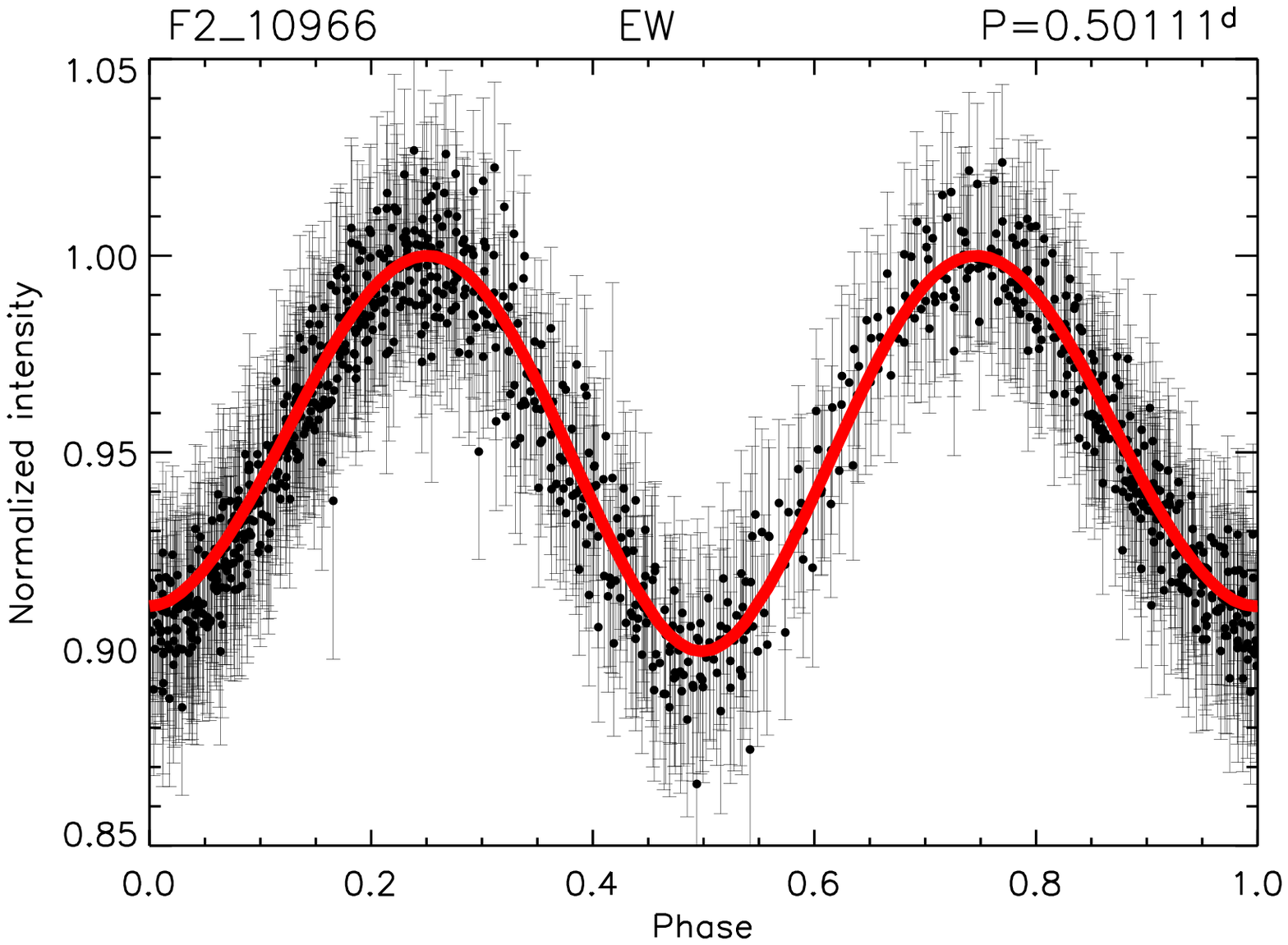}
\includegraphics[width=0.49\textwidth]{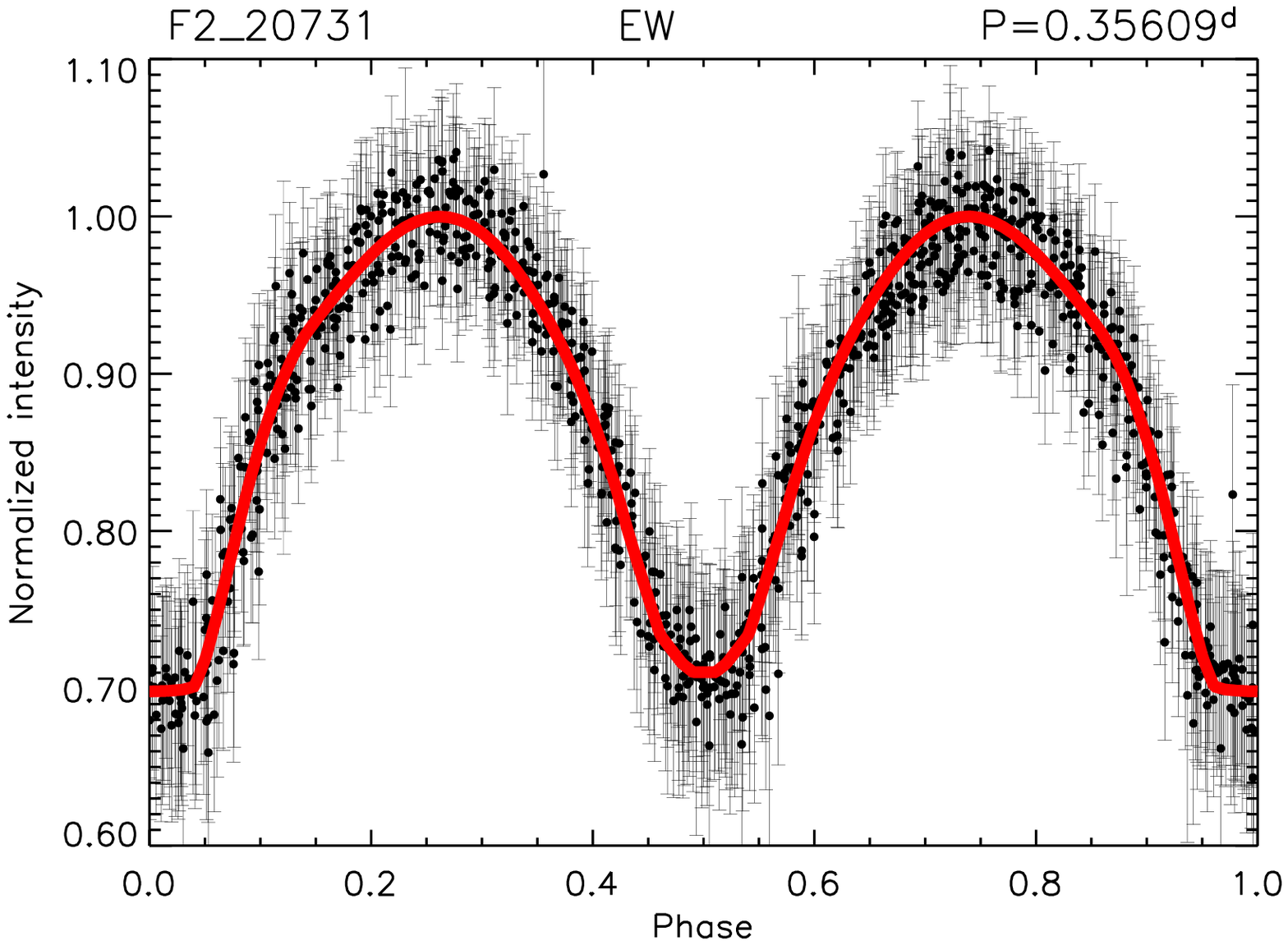}
\caption{Light curves of the modelled binary systems F2\_2633 (DV Per), F2\_04262 (NSVS 1926064), F2\_10496 (NSVS 1913469), F2\_10966 and F2\_20731 (NSVS 4052900). Points represent the measurements and the red solid line denotes the fit of the light curve solutions. See the electronic edition of the Journal for a color version of this figure.\label{fig:lcs_modelled}} %\label{fig:lcs_modelled2}
\end{figure*}

The automatic detection of periods through AoV during data reduction leads to a period value of $0.80714$\,days for DV~Per, which has the identifier F2\_02633 in our data set. This is in good agreement with the period from \citet{hoff44}, but there is no sign for a secondary occultation (Fig. \ref{fig:lcs_modelled}). The duration of the eclipse with this period would be $24$\,\% of the total orbital period. This is only possible, if both objects are quite close too each other. Usually, the light curves of such scenarios show ellipsoidal variations between the eclipse and occultation events, which are not present in our measured light curve. The modelling consequently inducts the physically not plausible solution of a mass less, dark secondary object which has the same size as the primary object and is almost attached. Due to these facts we decided to double the period of this system to $P=1.61427$\,days leading to a light curve with two separated minima of equal depth. This indicates the light curve of a detached binary system.

In order to model the light curve, we set the period to $1.61427$\,days and fixed the mass ratio at $q=1.0$ arbitrarily, which is reasonable since the two nearly equal minima suggest similar stars. Adjusted parameters were the fill-out factors of the two objects, the inclination, phase-shift and the temperature of the secondary. With the help of the 2MASS colors $J$ and $K$ we estimated the temperature of the primary star to be $6300$\,K and kept it fixed at this value. Remaining parameters were fixed according to Table \ref{tab:modelparameters}. The found solution suggests a physically plausible detached configuration (Fig. \ref{fig:lcs_modelled} and Table \ref{tab:modelparameters}).

A detailed spectroscopic study to measure the true mass ratio as well as multicolor photometry is highly desirable for a further, detailed study of this system. Combination with radial velocity measurements will clearly decide between the old and newly proposed period value. Furthermore, this detached system with two nearly equal stars in a short period orbit is suitable for testing stellar evolutionary models.

\clearpage

\subsubsection{EN Per}

EN Per (F2\_14415 in our data set) is a moderately faint ($V = 13.4$\,mag) \citep{samu09} Algol-type eclipsing binary with a relatively long orbital period of $10.24665$\,days, making it a potential candidate for an eccentric eclipsing binary. Such cases are suitable for testing theory of general relativity and stellar evolutionary theories, e.g. \citet{clar10}. We observed only the beginning of the primary eclipse once and the part of the ascending branch of the primary eclipse another time (see Fig. \ref{fig:lcs}, object F2\_14415). The center of the primary eclipse is not covered by our observations and thus we are not able to confirm the total amplitude of $0.9$\,mag given in VSX. No full secondary eclipse was observed by us, but the points at phase 0.465 seem to be part of the secondary eclipse. If this is confirmed by future observations it is an eccentric system. An out-of-eclipse variation with a peak-to-peak amplitude of $0.1$\,mag is present and seems to be periodic. However, we are not able to characterise it from our data set.

\subsubsection{NSVS 1926064}

The variability of the EW-type eclipsing binary star F2\_04262, also known as suspected variable star NSVS 1926064 \citep{gett06}, is now confirmed. We find an identical orbital period of $0.40747$\,days (Fig.~\ref{fig:lcs_modelled}). The maximum brightness reaches $11.5$\,mag. Fortunately, the system shows total eclipses making it suitable for modelling since total eclipses with $i \approx 90^\circ$ strongly reduce the degeneracy of possible light curve solutions \citep{ruci73}.

In order to find a light curve solution with the used model \citep{csiz09} we fix the temperature of the primary at $T_1=6000$\,K based on the period-color-relation of \citet{wang94}. Table~\ref{tab:modelparameters} shows an overview of the free and fixed parameters together with the results. Fig.~\ref{fig:lcs_modelled}. illustrates the light curve plus the determined fit. The system has a high inclination of $(88^\circ\pm 5)^\circ$, moderate third light of $13$\,\% coming from a non-resolved star, and a common mass ratio of $q=4.11\pm 0.11$. The system has a very high fill-out factor of $f=0.627$, showing that it is close to the phase when the two stars will merge, representing a probable late stage of their evolution.

The more massive component is the cooler one and is thus classified as a W-subtype system. Therefore it serves as an important target to understand the evolution and structure of low mass ratio contact binaries. Due to moderate brightness, F2\_04262 is a suitable target for future investigations, e.g. radial velocity measurements, spectral type determination, period investigation and stellar spot studies.

\subsubsection{NSVS 1913469}

The object NSVS 1913469 (F2\_10496) is declared as EW in VSX with an amplitude of $0.31$\,mag in R1 and a period of $0.42167068$\,d. We confirm  the variability type and found a comparable period of $0.42166$\,d but a smaller amplitude of $0.19$\,mag (Fig.~\ref{fig:lcs_modelled}). The light curve shows the O'Connell-effect, i.e. the secondary maximum is a bit higher than the primary. In well-studied cases it was found that EW-type variables show cycle-to-cycle variations due to variable spot sizes, temperatures and distribution \citep{csiz04b}. For further investigation of this effect multicolor-photometry is needed. The system belongs to the A-subtype class of contact binaries.

There is no clear sign of a total eclipse, suggesting that the object is probably a medium-inclination system. Via light curve modelling we found an inclination of $\sim75^\circ$. It is a low-degree-of-contact system with a small fill-out factor of $f=0.036$. The remarkable amount of third light is $L3/(L_1+L_2+L_3) \approx 0.48$, which may originate from a nearby, not fully resolved optical companion. This high value of the third light squeezes the amplitude to the observed value and can explain the large difference in the observed amplitudes. Similary high third light values are common in other EW-binaries, e.g. with a record of  $L3/(L_1+L_2+L_3) \approx 0.93$ in V758~Cen \citep{lipa85, csiz04a}. The results for NSVS 1913469 are shown in Table~\ref{tab:modelparameters} and in Fig.~\ref{fig:lcs_modelled}.

\subsubsection{NSVS 4052900}

This EW-type eclipsing binary star F2\_20731 belongs to the A-subtype. It has a low fill-out factor of $12.3$\,\% and significant third light of $L_3/(L_1+L_2) = 23$\,\%. This light probably comes from a nearby companion star, which is not fully resolved with our pixel scale. The light curve shows total primary eclipse, which is confirmed by the light curve modelling result ($i\approx85^\circ$). Although its high inclination makes it interesting for further studies, it is moderately faint with $13.2$\, mag at maximum brightness. The modelled light curve can be found in Fig. \ref{fig:lcs_modelled} and the parameters are listed in Table \ref{tab:modelparameters}.

The system was marked as a suspected variable in VSX and is listed under the name NSVS~4052900. We confirm the variability type EW and the period of $0.35609$\,d. However, the measured amplitude is $0.4$\,mag and differs by a factor of 2.3. The reason can be different third light due to different pixel scales of the telescopes used. Photometric follow-up with higher angular resolution is needed in order to determine the amplitude with higher precision.

\subsubsection{Remaining Suspected Variables}

The VSX denotes for object NSVS 1910955 (F2\_13911) a period of $1.73170386$\,d and an amplitude of $0.81$\,mag in R1 for this EA-type variable star. Period and type are confirmed, but the determined amplitude differs by a factor of 4.3, since we found an amplitude of $0.19$\,mag.

VSX denotes variability type DSCT and a period of $0.1441$\,d for the object VSX~J024717.4+504506 (F2\_24176). The amplitude is not listed there. We can confirm the variability type DSCT. Our determined period is larger ($P= 0.14908$\,d) and we measure an amplitude of $0.08$\,mag. 

The object VSX J023706.8+503557 (F2\_25347) is declared as a DSCT with $0.0865$\,d period. We found the same variability type DSCT but a different period of $0.09034$\,d.

\subsection{Newly Detected Extrinsic Variables}\label{sec:extvars}

The majority of the newly detected variable stars are eclipsing binaries with in total 69 detected objects. The largest subgroup of the binaries are the EW-type stars with 42 new detections. Stars of the EA sub-type amount to 21 new detections and the smallest subgroup are the semi-detached binaries of sub type EB with 6 representatives.

\begin{figure*}
\caption{Phase folded light curves of the detected variable stars. The other light curves are available in the electronic edition\label{fig:lcs}}
\centering
\includegraphics[width=0.237500\textwidth]{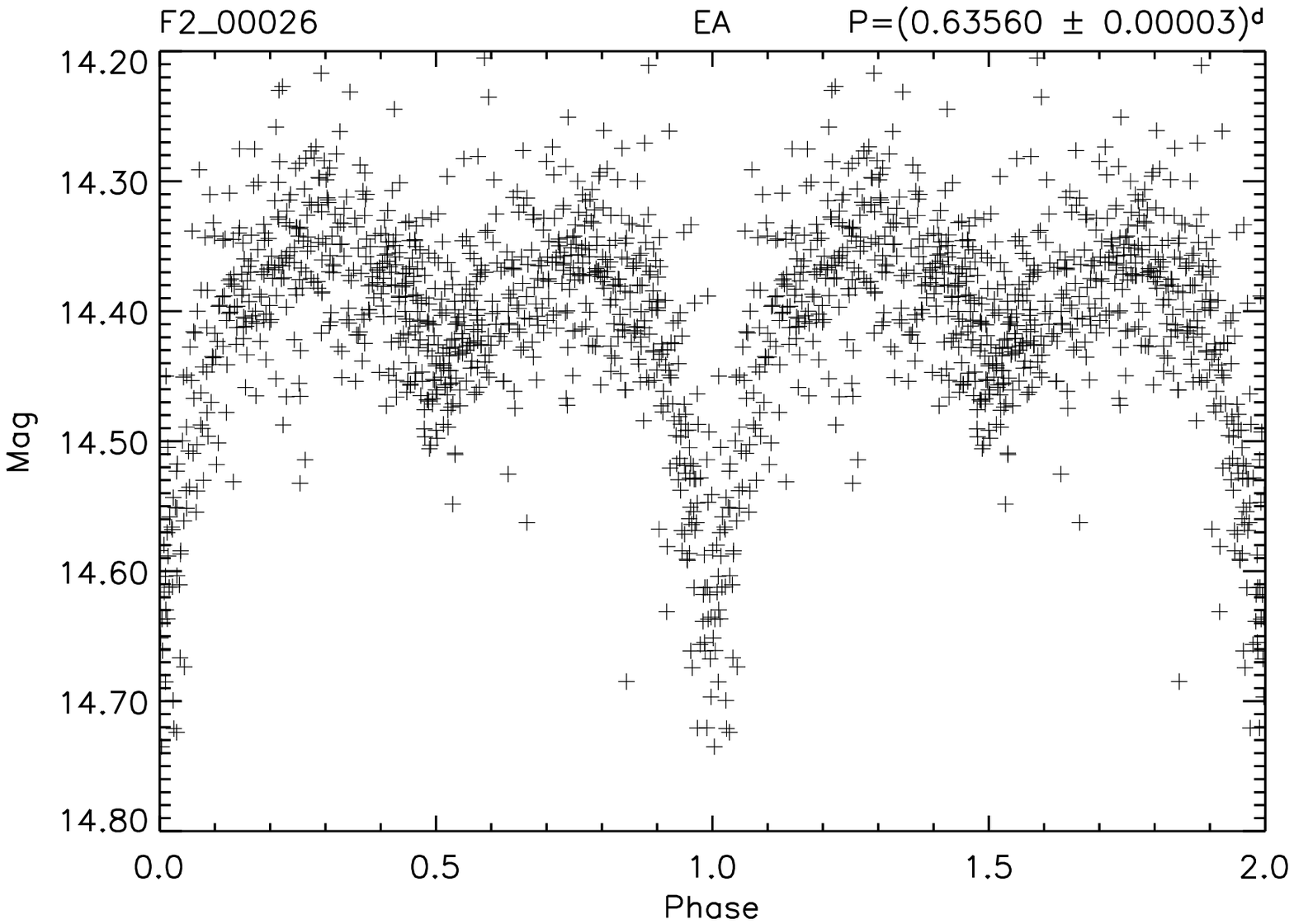}
\includegraphics[width=0.237500\textwidth]{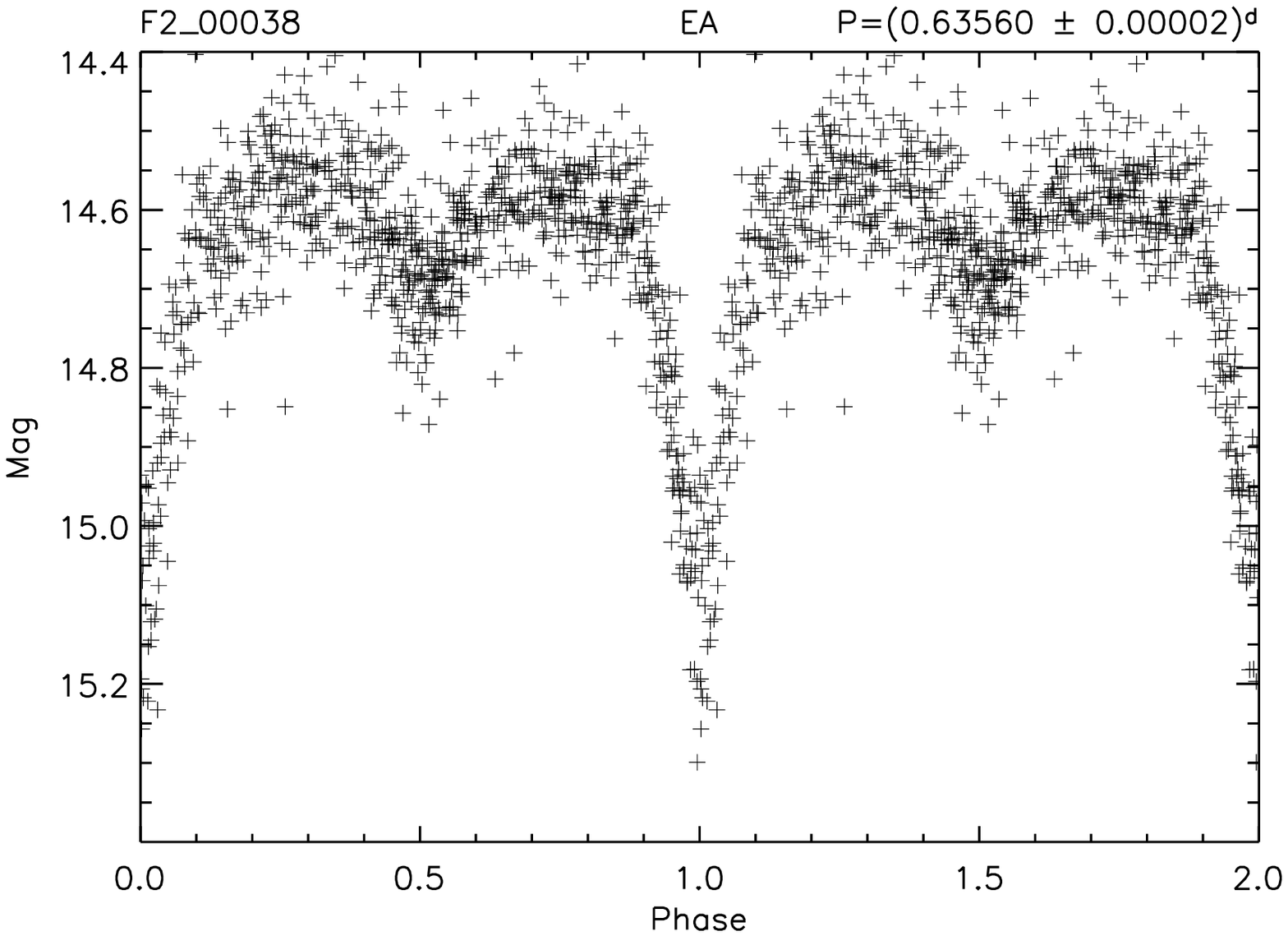}
\includegraphics[width=0.237500\textwidth]{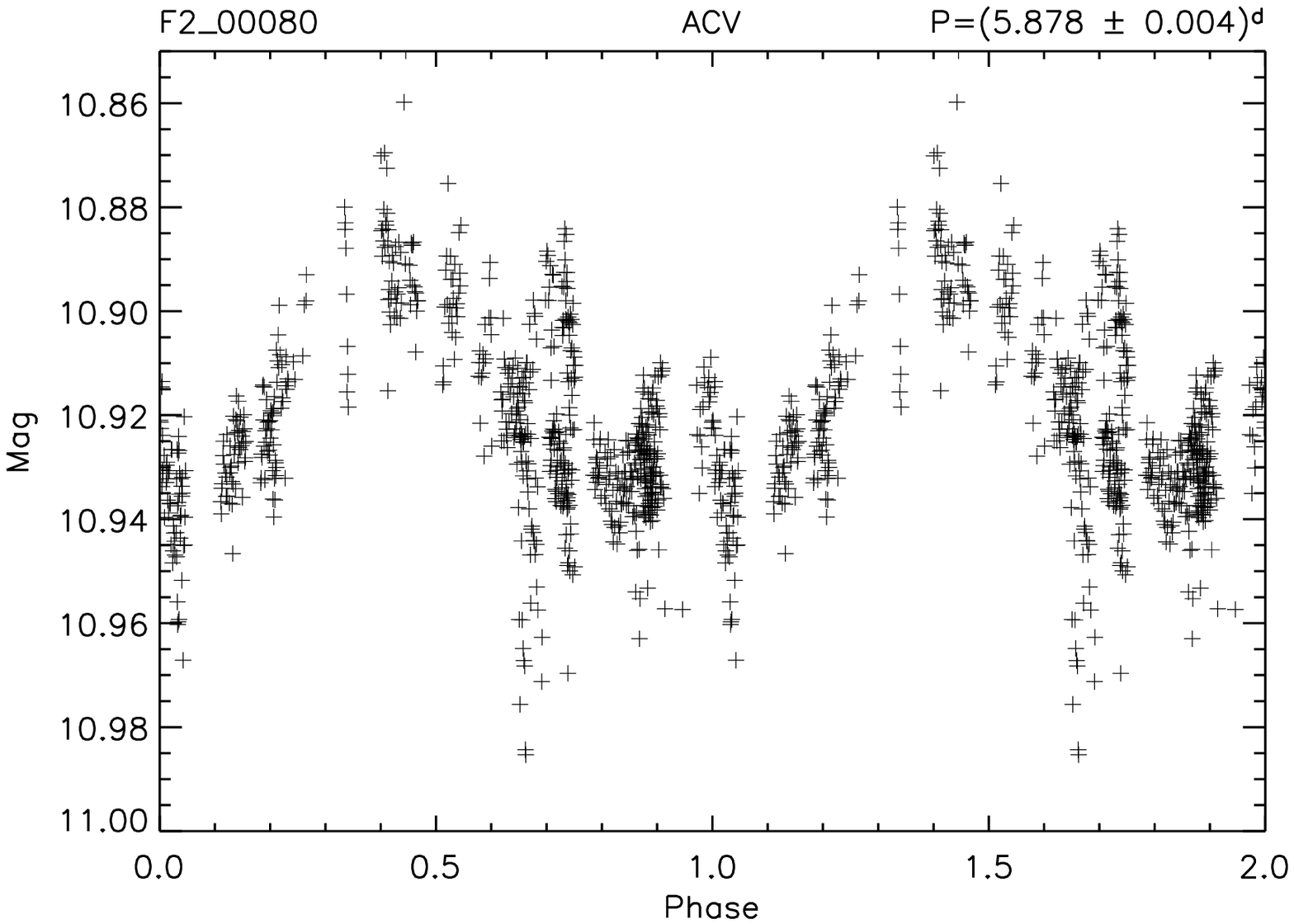}
\includegraphics[width=0.237500\textwidth]{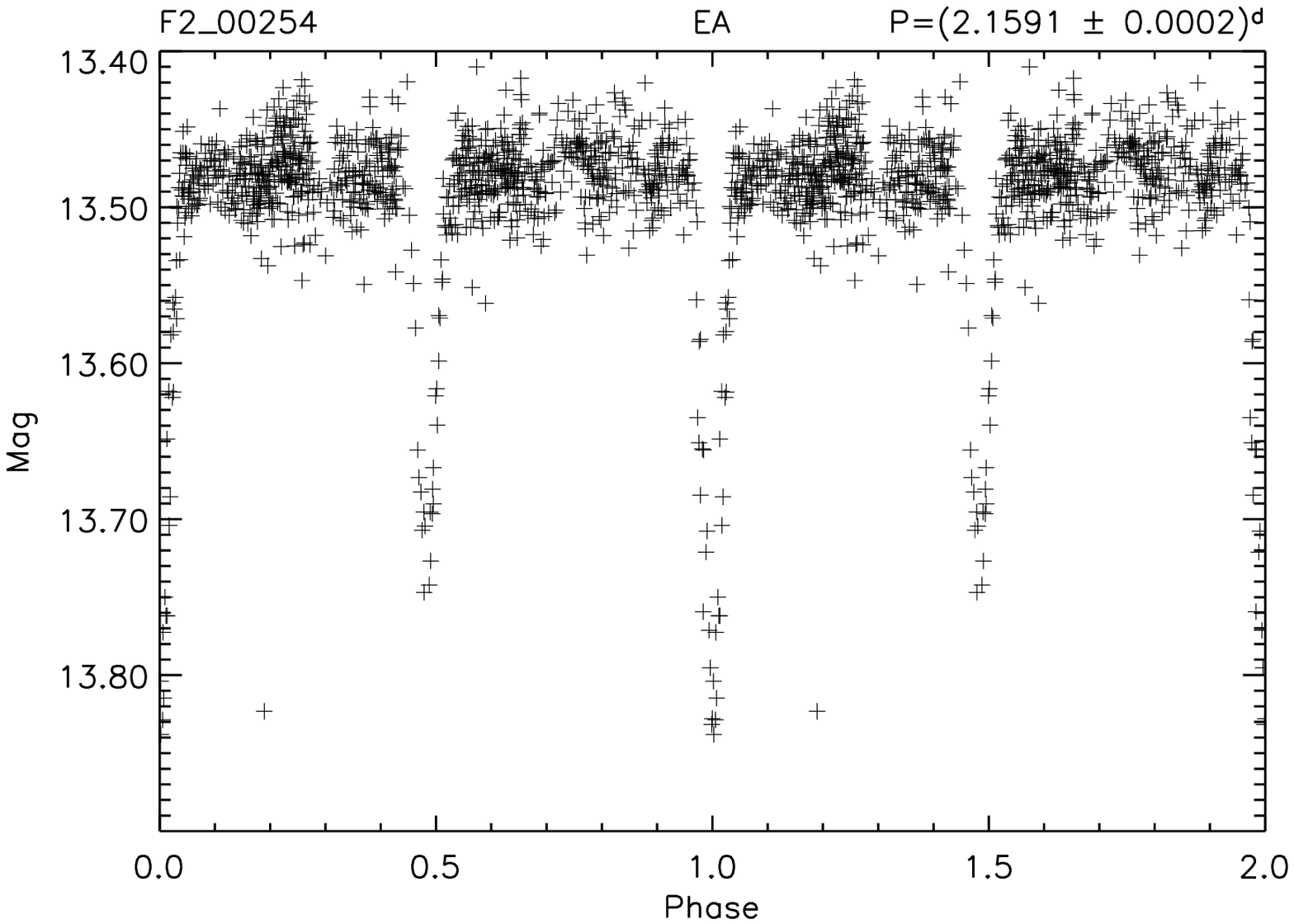}
\includegraphics[width=0.237500\textwidth]{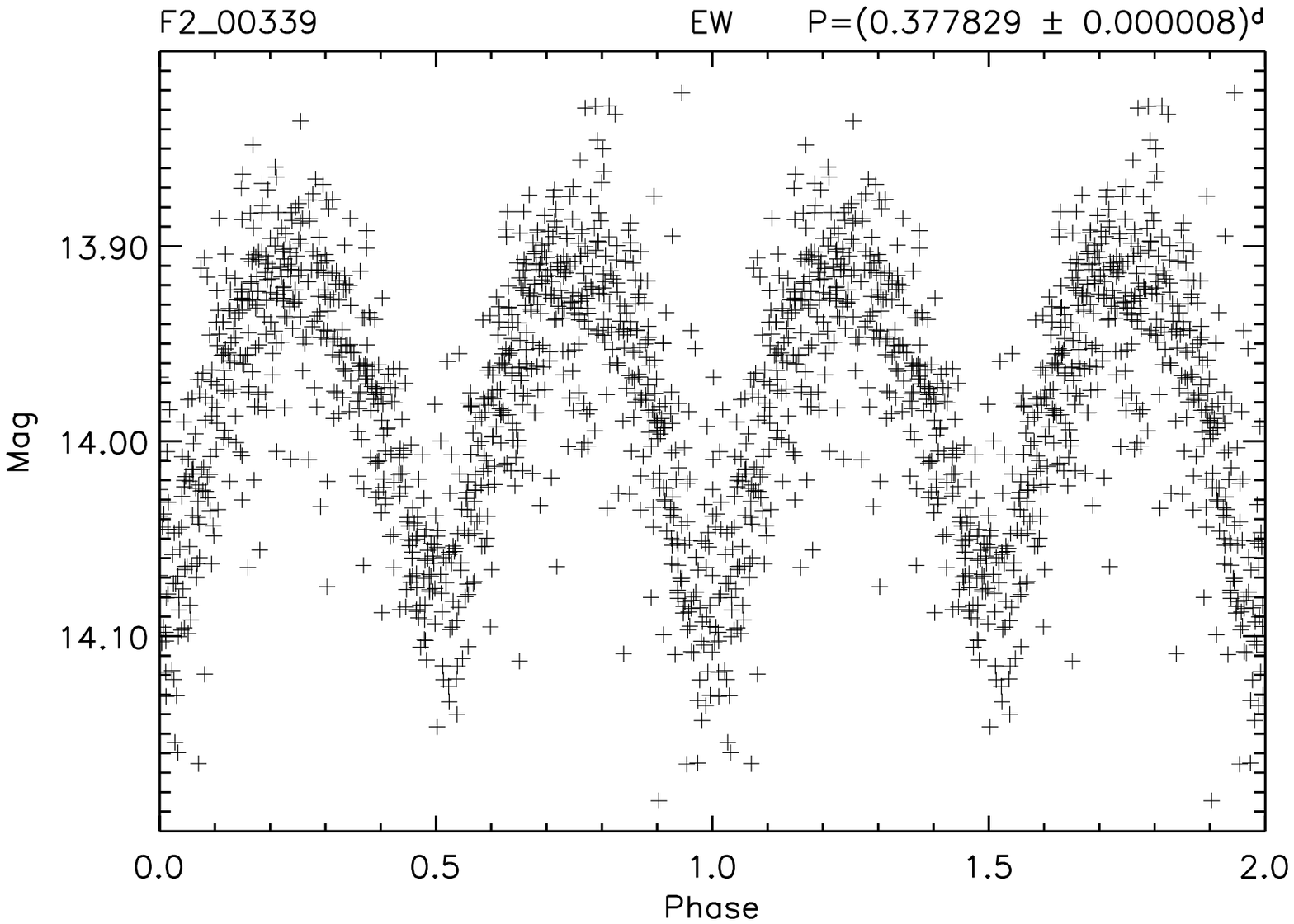}
\includegraphics[width=0.237500\textwidth]{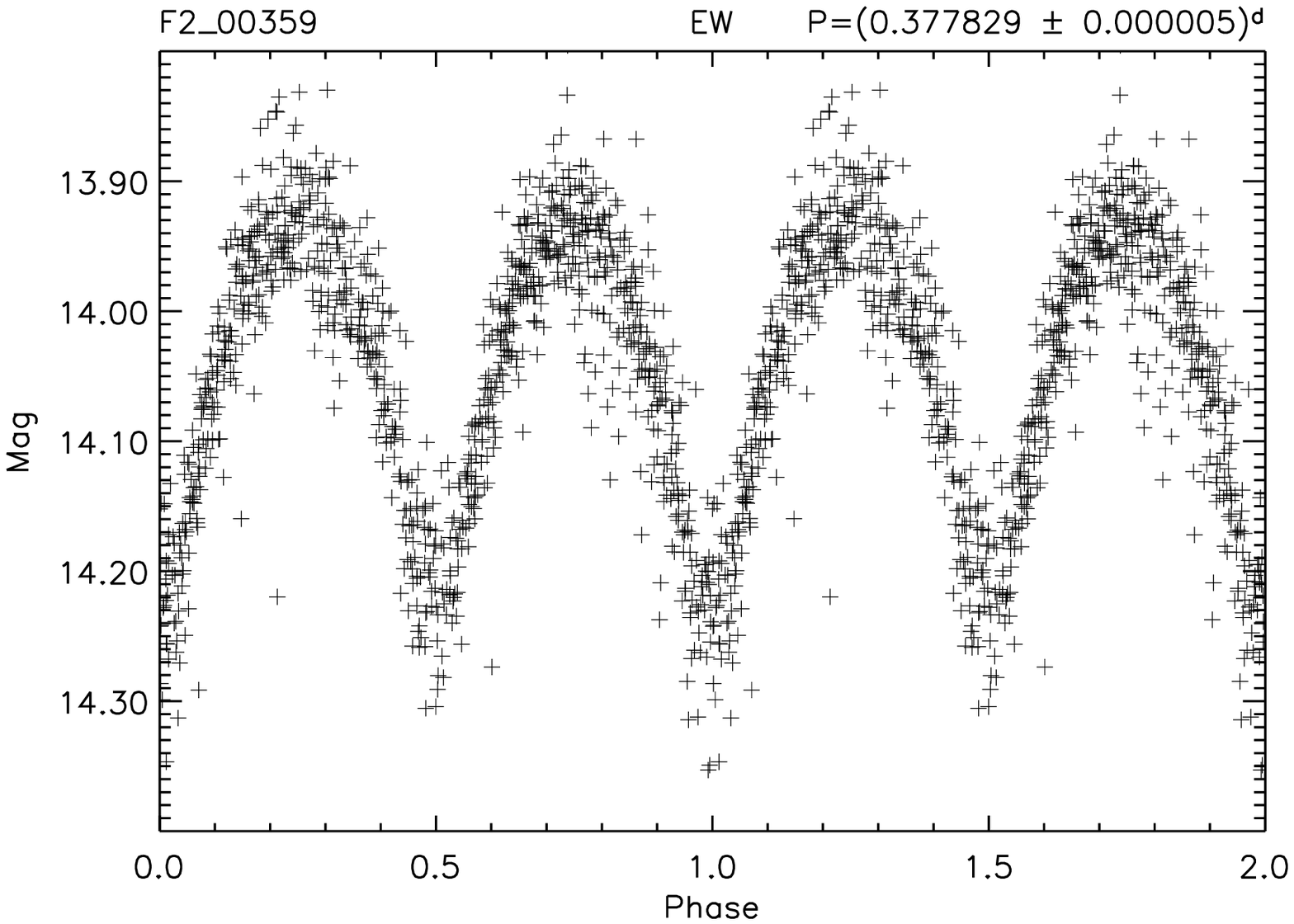}
\includegraphics[width=0.237500\textwidth]{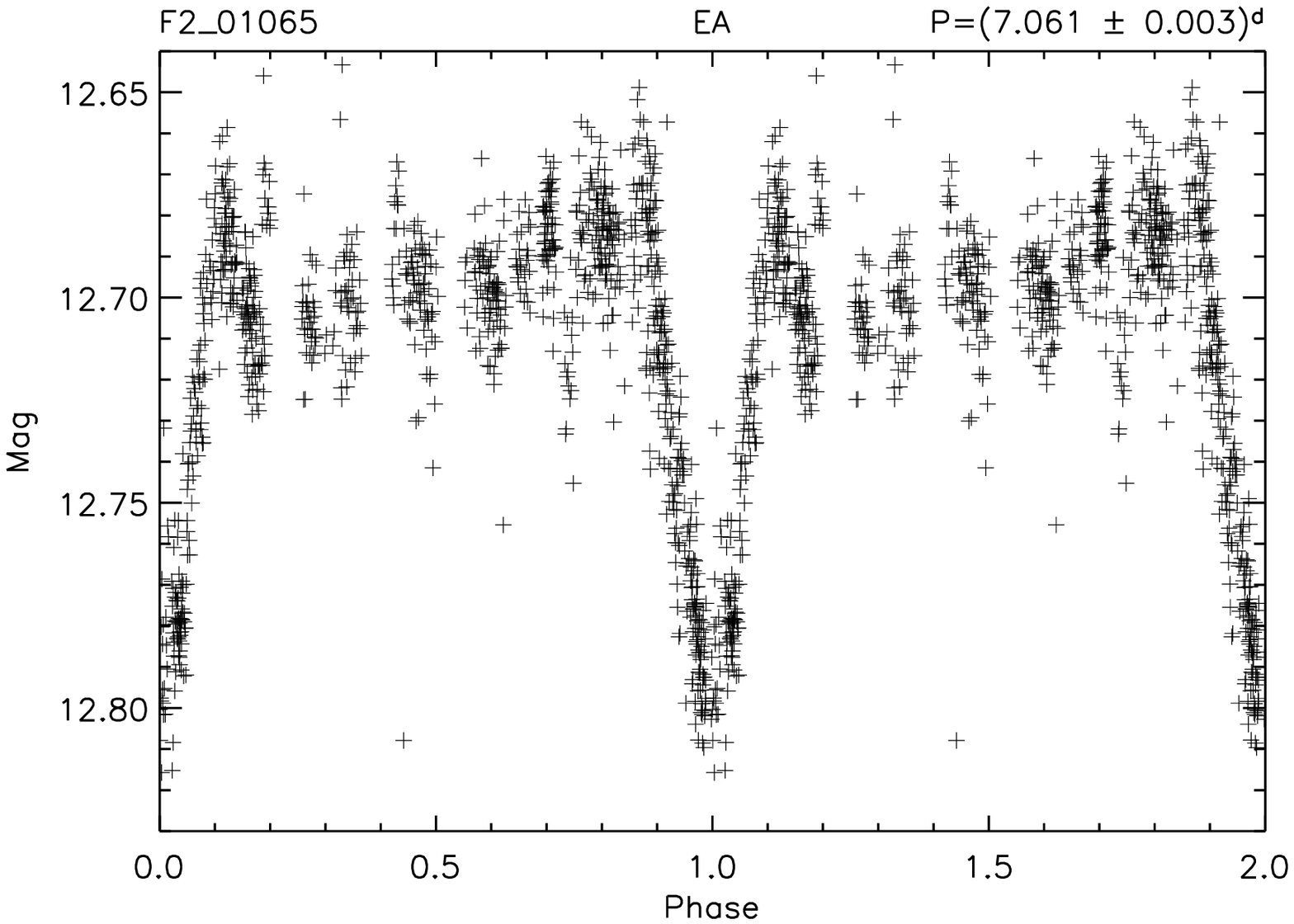}
\includegraphics[width=0.237500\textwidth]{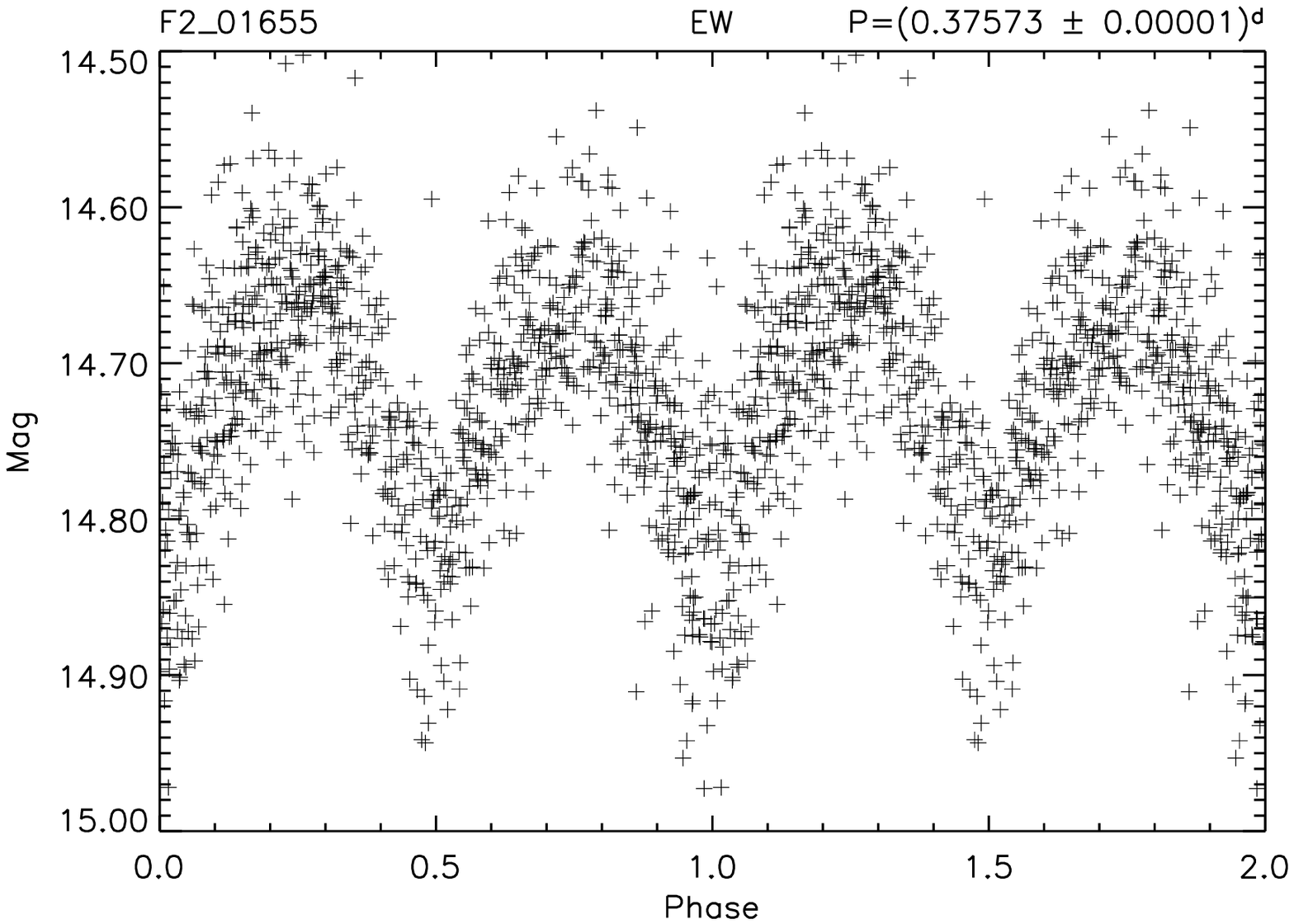}
\end{figure*}

We found 11 newly detected rotating variable stars, split into 5 ACV type, 5 ELL type and 1 spotted variable star. Details to these objects can be found in Table \ref{tab:varcat} and Fig. \ref{fig:lcs}. 

Below, some cases of special interest are discussed in detail. Among these objects, we want to emphasize the eccentric binaries F2\_00254 and F2\_03278. Eccentric eclipsing binaries are key objects to critically check the stellar structure and evolutionary models via the extremely sensitive 'apsidal-motion test' \citep{bohm92}. Only 18 adequate systems have been available in a recent study \citep{clar10}. Discoveries of new eccentric eclipsing binaries, like this present cases, help to extend the sample with new possible targets.

\subsubsection{F2\_00254 -  an eccentric binary}

The moderately bright ($R=13.48$\,mag) object F2\_00254 is a newly discovered Algol-type variable with a low amplitude of $0.3$\,mag. The secondary eclipse is displaced to phase $\varphi=0.48$ indicating an eccentric orbit, although it has a relatively short orbital period for eccentric binaries of $2.16$\,days. From the displacement of the secondary and the lengths of the eclipses one can suspect a small eccentricity of $0.03$ with an argument of periastron of $36^\circ$. A low number of measurements during the eclipses makes it an unsuitable object for light curve modelling. For further investigation the system requires spectroscopic and radial velocity observations as well as precise timings.

\subsubsection{F2\_03278 - another eccentric binary}

The object F2\_03278 stands out due to the large displaced secondary eclipse at $\varphi=0.36$. Despite its relatively short period of $1.85140$\,d it has a large eccentricity of $0.24$ with an argument of periastron of $16^\circ$. The total variation does not exceed $0.17$\,mag. It is a bright object with $R=12.7$\,mag and is therefore an appropriate target for photometric and spectroscopic follow-up observations.

\subsubsection{F2\_08747 and F2\_08777}

One of these two stars is a short-period Algol-object. They are separated by 18~arcseconds and due to limits in spatial resolution the light curves are affecting each other. Thus, it is not clear which of these two stars is variable. Further studies with higher angular resolution are needed in order to distinguish between them. We do not carry out a light curve modelling of this system since our observational data is not complete enough to decide for the correct period.

The real variable among these two objects has a considerably short variability period. Eclipses occur in every 0.54 days, while most of the Algol-type variables have much larger orbital periods. The duration of the eclipse is 25\% of the period, extremely long. This may imply that the orbital period is twice the value published here in reality. If the given period is true, it is a quite important object, because EA-type binaries in this period regime are rare \citep{pacz06}.

\subsubsection{F2\_10966}

This EW-type binary is largely contaminated by a nearby star, not fully resolved by our pixel scale (Fig.~\ref{fig:lcs_modelled}). The third light was found to be large with $L_3/(L_1+L_2) =  0.55$. The system has a low inclination of $\sim50^\circ$ and as \citet{ruci73} pointed out the determination of mass ratio by photometry in the case of low-inclination contact binaries is quite difficult. This fact and the low quality photometry caused by the nearby disturbing companion make the results uncertain. For instance the mass ratio can be constrained only to be $q=3.66\pm0.59$. The larger star is the hotter one, so the system belongs to the A-subtype of the contact binaries. Results of the light curve modelling are given in Table~\ref{tab:modelparameters} and Fig.~\ref{fig:lcs_modelled}.

\subsubsection{F2\_12012}

The interesting feature of this Algol-type variable is the linearly decreasing brightness between primary and secondary minimum, followed by a new increase from secondary to primary. This indicates a relatively strong reflection effect. The system is also slightly eccentric because the secondary is at phase $\varphi = 0.502$, which is unusual for such a short period binary ($P=1.25$ days).

\subsubsection{F2\_20412}

This object is a high amplitude EA-type star which shows variations of $1.01$\,mag. Its period of $1.96$\,days is close to two days and possibly explains, why this variability could escape from discovery until today despite its large amplitude.

It is conspicuous, that the few observations of the primary minimum are not matching each other. This may suggest some period variation, which is a common effect in semi-detached classical Algol-system. The light curve also suggests, that there can be out-of-eclipse variations caused either by pulsation or gas-streams in the system. The system evidently requires more photometric studies.

\subsubsection{F2\_21539}

The $\beta$ Lyrae-type eclipsing variable F2\_21539 has an orbital period close to one day ($P\approx 0.903$ days), that is why the second half of its light curve is better sampled than the first one. With only 20\% of amplitude difference between primary and secondary it seems to be a low-inclination system. The asymmetric light curve can be caused by stellar spots and/or discs. The system shows considerable light curve shape variations in different periods. EB-type variables often exhibit strong period- and light curve variations making it unsuitable for light curve modelling and hence it is not carried out here.

\subsection{Newly Detected Intrinsic Variables}\label{sec:intvars}

The second largest group are the pulsating stars with 39 detected objects, out of which 37 are new detections. The subgroup of DSCT-stars is most dominant here with 33 new detections. Remaining objects are one Cepheid and three RR Lyrae-type stars

The most intriguing member of the intrinsic variables is the object F2\_09743, which is a typical mono-periodic $\delta$ Scuti-type star. With a short period of $0.18$\,days, a high amplitude of $0.24$\,mag and a brightness of $12.85$\,mag this object is a favourable target for future studies. Note, that most of the DSCT stars show some light curve modulations, while this object featured a very stable light curve during the 5 years of observations between 2001-2006.

\section{Summary} \label{sec:summary}

We observed with BEST a $3.1^\circ\times 3.1^\circ$ field in the constellation Perseus in \nightstotal nights between 2001 August and 2006 December from the locations TLS and OHP. In a sample of \starstotal light curves we detected \varstotal variable objects, out of which \varsnew are new detections.

Variability of the known variables DV, EF, EN, EE, EH, V670, V726, and V727 Persei is confirmed. We present an updated period value and the first light curve parameter estimation for DV Per, which seems to be a suitable object for stellar evolutionary studies. The period for EN Per is confirmed. For the remaining 6 known variables we are just able to confirm their variability. Statements about type, period, and amplitude are not possible due to the poor phase coverage.

Furthermore, the variability of 12 previously suspected variables is confirmed. For six of them we are able to confirm the type but find slightly different periods and occasionally totally different amplitudes of variation. The remaining six cases are confirmed as clearly variable objects, but we are not able to determine periods for this ones due to long- or semi-periodicity. For 4 stars with known variability the system parameters were calculated by modelling the light curve for the first time.

Among the \varsnew new detected variables 36 new pulsating variables, 69 new eclipsing binaries and 12 other extrinsic variable stars were found. The percentage of variable objects in the BEST data set is $0.45$\,\% and thus comparable to other photometric wide-field surveys.

\acknowledgments

\textit{Acknowledgements} We thank the Observatoire de Haute Provence for great support of the BEST survey. This publication makes use of data products SIMBAD, 2MASS and USNO-A2.0 as well as the AAVSO variable star index. We are also grateful to the observers Michael Weiler, Tino Wiese, Susanne Hoffmann, Christopher Carl and Martin Dentel.

\begin{deluxetable}{lcccccccccc}
\tabletypesize{\scriptsize}
\tablewidth{0pt}
\rotate
\tablecaption{Variable Stars detected in the observed BEST field with coordinates, magnitude, period, epoch, amplitude, and variability type. Stars affected by crowding are flagged with c. The flag k denotes previously known objects and their ID from VSX or GCVS can be found in the last column. The Epoch $T_0$ is given in reduced Julian date [rJD] in respect to $T=2,450,000.0$. It denotes the first minimum in the light curve.\label{tab:varcat} (This table is available in its entirety in machine-readable and Virtual observatory (VO) forms in the online journal. A portion is shown here for guidance regarding its form and content.)}
\tablehead{ %
 \colhead{BEST ID} %
&\colhead{flag} %
&\colhead{2MASS ID}  %
&\colhead{$\alpha(J2000.0)$}  %
&\colhead{$\delta(J2000.0)$}  %
&\colhead{R$_B$ [mag]}  %
&\colhead{T$_0$ [rJD]}%
&\colhead{P [d]}    %
&\colhead{A [mag]}  %
&\colhead{Type}  %
&\colhead{Other names} %
}
\startdata

F2\_00026  &c   &02304181+5329285  &$02^h30^m42.0^s$  &$ 53^\circ29'28.9"$  &14.40  &2196.541  &    $0.63560 \pm 0.00003$  &     $0.25 \pm 0.09$  &EA  &\\
F2\_00038  &c   &02304450+5329209  &$02^h30^m44.5^s$  &$ 53^\circ29'20.5"$  &14.63  &2196.538  &    $0.63560 \pm 0.00002$  &       $0.5 \pm 0.1$  &EA  &\\
F2\_00080  &    &02294352+5328354  &$02^h29^m43.5^s$  &$ 53^\circ28'35.4"$  &10.93  &2280.534  &        $5.878 \pm 0.004$  &       $0.1 \pm 0.2$  &ACV  &\\
F2\_00254  &    &02340421+5328480  &$02^h34^m04.2^s$  &$ 53^\circ28'47.8"$  &13.48  &2197.066  &      $2.1591 \pm 0.0002$  &     $0.19 \pm 0.05$  &EA  &\\
F2\_00339  &c   &02432992+5329386  &$02^h43^m30.0^s$  &$ 53^\circ29'39.5"$  &13.97  &2196.736  &  $0.377829 \pm 0.000008$  &     $0.17 \pm 0.05$  &EW  &\\
F2\_00359  &c   &02433226+5329253  &$02^h43^m32.3^s$  &$ 53^\circ29'26.3"$  &14.02  &2196.735  &  $0.377829 \pm 0.000005$  &     $0.31 \pm 0.05$  &EW  &\\
F2\_01065  &    &02364668+5322516  &$02^h36^m46.6^s$  &$ 53^\circ22'53.1"$  &12.70  &2197.098  &        $7.061 \pm 0.003$  &     $0.11 \pm 0.03$  &EA  &\\
F2\_01655  &c   &02450572+5319123  &$02^h45^m05.6^s$  &$ 53^\circ19'10.9"$  &14.72  &2195.597  &    $0.37573 \pm 0.00001$  &     $0.17 \pm 0.06$  &EW  &\\
\enddata
\end{deluxetable}

\end{document}